\documentclass[preprint]{elsarticle}

\usepackage{lineno,hyperref}
\usepackage{amsfonts}
\usepackage{amsmath}
\usepackage{algorithm}
\usepackage{algorithmic}
\usepackage{makecell}
\usepackage{multirow}
\modulolinenumbers[5]

\begin{document}

\begin{frontmatter}

\title{TFDPM: Attack detection for cyber-physical systems with diffusion probabilistic models}

% Group authors per affiliation:
\author{Tijin Yan$^{a}$}
\author{Tong Zhou$^{a}$}
\author{Yufeng Zhan$^{a}$}
\author{Yuanqing Xia$^{a,*}${\corref{mycorrespondingauthor}}}
\address[mymainaddress]{Key Laboratory of Intelligent Control Decision of Complex Systems, School of Automation, Beijing Institute of Technology, Beijing 100081, P. R. China.}
\cortext[mycorrespondingauthor]{Corresponding author: Yuanqing Xia}
\ead{xia\_yuanqing@bit.edu.cn}
%\author{XXXXXXXXXXXX\fnref{myfootnote}}

\begin{abstract}
%TODO: 任务+现有方法，我们提出，此外，实验证明
%With the development of AIoT, data driven attack detection methods for cyber-physical systems (CPSs) have attracted lots of attention. However, existing methods usually adopt tractable distributions to approximate data distributions, which are not suitable for complex systems. Besides, the correlation of the data in different channels does not attract sufficient attention. To address these issues, we propose TFDPM, a general framework for attack detection tasks in CPSs. Temporal pattern and feature pattern are firstly extracted. In particular, graph attention networks are used to explicitly model the correlation of the data in different channels. The extracted features are sent to a conditional diffusion probabilistic model, which is less restrictive on functional forms of the data distribution. Attacks are detected based on the difference between predicted values and observed values. In addition, to realize real-time detection , a conditional noise scheduling network is proposed to accelerate prediction process. We perform extensive experiments on three real-world CPS datasets and the results show that TFDPM outperforms existing state-of-the--art attack detection methods. The noise scheduling network is efficient for acceleration.
%use graph attention networks to explicitly model the correlation of the data in different channels.

With the development of AIoT, data-driven attack detection methods for cyber-physical systems (CPSs) have attracted lots of attention. However, existing methods usually adopt tractable distributions to approximate data distributions, which are not suitable for complex systems. Besides, the correlation of the data in different channels does not attract sufficient attention. To address these issues, we use energy-based generative models,  which are less restrictive on functional forms of the data distribution. In addition, graph neural networks are used to explicitly model the correlation of the data in different channels. In the end, we propose TFDPM, a general framework for attack detection tasks in CPSs. It simultaneously extracts temporal pattern and feature pattern given the historical data. Then extract features are sent to a conditional diffusion probabilistic model. Predicted values can be obtained with the conditional generative network and attacks are detected based on the difference between predicted values and observed values. In addition, to realize real-time detection, a conditional noise scheduling network is proposed to accelerate the prediction process. Experimental results show that TFDPM outperforms existing state-of-the-art attack detection methods. The noise scheduling network increases the detection speed by three times.

%This template helps you to create a properly formatted \LaTeX\ manuscript.
\end{abstract}

\begin{keyword}
attack detection, cyber-physical systems, energy-based models, graph neural networks
%\texttt{elsarticle.cls}\sep \LaTeX\sep Elsevier \sep template
%\MSC[2010] 00-01\sep  99-00
\end{keyword}

\end{frontmatter}

%\linenumbers

\section{Introduction}
%TODO: 背景，任务的定义； 现有方法；现有方法面临的挑战；针对此挑战，做什么改进；实验部分的结论。最后总结
% 信息物理系统定义，常用于监控。。通过
% 研究很久，现有方法很多；例如传统方法有，，，最近，基于深度学习的方法最近成为一种趋势。例如，基于预测的方式通过比较预测值与实际观测值的差异，选定合适的阈值来比较差异；基于重构的方式通过重构值与输入的差距来判断异。 此外，概率模型方法相比确定性方法
%Cyber-physical systems are computer systems where physical processes are monitored or controlled by softwares \cite{wu2018secure}. As a interdisciplinary subject, they have attracted lots of attention and they are commonly used for monitoring and controlling of industrial processes. It's very important to monitor

Cyber-physical systems (CPSs) are computer systems that monitor and control physical processes with software \cite{wu2018secure}. They integrate computing, communication, control and physical components. Due to potential security threats such as component performance degradation, human factors, network attacks, etc., their security and reliability have attracted lots of attention \cite{jiang2017recursive}. With the increasing complexity of modern CPSs, traditional rule-based attack detection methods cannot meet the requirements gradually. Recently, with the development of AIoT (AI + IoT), it has become a trend to build attack detection frameworks of CPSs based on data-driven machine learning algorithms.

Attack detection for CPSs has been an active research topic for a long time \cite{chalapathy2019deep,ahmed2016survey} and many methods have been proposed. For example, traditional unsupervised attack detection methods like isolation forest \cite{liu2008isolation} and DBSCAN\cite{schubert2017dbscan} have been successfully applied to industrial production. Recently, deep learning-based models have achieved great performance. For example, prediction-based methods like recurrent neural networks \cite{hundman2018detecting,salinas2020deepar,czarnowski2020deepfactors} use features extracted from historical data to predict future values. Then we can detect attacks based on the difference between the predicted values and actual observations. Reconstruction-based methods like autoencoders \cite{xu2018unsupervised} detect attacks according to the difference between reconstructed values and inputs. In addition, probabilistic generative methods which model the distribution of data are demonstrated more effective than deterministic methods \cite{su2019robust}. To some extent, existing methods have achieved good results on various attack detection tasks in CPSs.

Despite their success, how to fully and effectively extract features of the data is still a challenge. For a set of time series data collected from CPSs, the data of different channels are usually correlated with each other. However, it's difficult to obtain explicit formulas for modeling the correlation of different dimensions. Methods based on recurrent neural networks \cite{wen2017multi} or attention \cite{qin2017dual} usually do not explicitly model the correlation of the data in different dimensions, which limits the prediction performance of the model. 

In addition, learning the distribution of data is also extremely challenging. Generally, there is no explicit form for the distribution and the amount of data is limited. Therefore, the maximum likelihood estimation method cannot be directly used for distribution estimation. Existing methods usually employ tractable distributions to approximate the distribution of data, which set the model with a certain form \cite{song2021train}. For example, flow-based methods \cite{kobyzev2020normalizing} model the data distribution by constructing strictly invertible transformations. In addition, the data need to be modeled as a directed latent-variable model in variational autoencoders. The assumption of the data distribution makes these models easier to optimize. However, tractable distributions are not always suitable for modeling data distributions.

To address these problems, we propose TFDPM, a general framework based on conditional energy-based generative models. Firstly, TFDPM simultaneously extracts temporal pattern and feature pattern from historical data.  In particular, we use graph neural networks to explicitly model the correlation of the data in different channels. Then the extracted features are served as the condition of a conditional probabilistic generative model. In this paper, we use energy-based models (EBMs) because they are much less restrictive in functional forms and have no assumptions on the forms of data distribution compared with other generative models \cite{song2021train}. The predicted values can be obtained through the conditional generation network. What's more, in order to meet the requirements of real-time detection, an extra conditional noise scheduling network is proposed to accelerate the prediction process. The experimental results on three real-world datasets show that our method achieves significant improvement compared to all baseline methods. In addition, the noise scheduling network increases the detection speed by three times.

In summary, the contributions of this paper can be summarized as 

\begin{enumerate}
	\item We propose TFDPM, a general framework for anomaly detection for CPSs. The network consists of two modules at each time step: temporal pattern and feature pattern extraction module and conditional diffusion probabilistic generative module. 
	\item To the best of our knowledge, TFDPM is the first model that uses diffusion-based probabilistic models on anomaly detection tasks for CPSs.
	\item In order to meet the requirements of online real-time detection, a conditional noise scheduling network is proposed to accelerate the prediction process of TFDPM.
	\item Experimental results demonstrate that our method outperforms other latest models on three real-world CPS datasets. And the noise scheduling network increases the detection speed by three times.
\end{enumerate}

The paper is organized as follows. Some related works are introduced in Section \ref{relatedwork}. Then some preliminaries of TFDPM are shown in Section \ref{preliminaries}. In Section \ref{methods}, the network architecture, training process and prediction process of TFDPM are introduced. The conditional noise scheduling network is also introduced. Then we perform extensive experiments and results and analysis are presented in Section \ref{experiments}. In the end, conclusions and future works can be found in Section \ref{conclusion}.
% 如何有效提取数据特征的研究仍不充分。此外，建模数据分布也具有很大的挑战性，

% 为了解决上述的问题，我们提出了基于。。。的模型，采用gat显式建模数据维度的相关性，此外，采用对函数约束较少的基于能量的模型。此外，为解决实时性问题，xxx

% 实验表明在xxx数据集上取得了xx效果，特别的，参数和模块的重要性。

%总结：

% 各部分之间的安排

\section{Related work}\label{relatedwork}
In this section, methods on attack detection tasks for CPSs are firstly introduced. Then energy-based generative models are presented. In the end, we briefly introduce graph neural networks.

\subsection{Attack detection for CPSs.}
Many works have been proposed for data-driven attack detection tasks in recent years and can be classified into three categories. The first category of approaches is machine learning-based methods (\cite{ding2015yading,huang2016time,li2018robust}). These methods usually divide time series into many segments and then cluster data based on distance, such as dynamic time warping and shape-based distance. Outliers of clusters are considered anomalies. The selection of similarity measurement is very important for clustering-based methods. The second category of approaches is reconstruction-based methods (\cite{malhotra2016lstm,mirsky2018kitsune,li2019mad,su2019robust}). These methods usually construct an auto-encoder network, which model the distribution of the entire time series and reconstruct the original input based on latent representations. Attacks are detected by the reconstruction probability (\cite{an2015variational}) or the difference between reconstruction values and real values. The dimension of latent variables is very important for auto-encoder networks. The third category of approaches is forecasting-based methods (\cite{hundman2018detecting,zong2018deep,yan2021scoregrad}), which detect attacks based on prediction errors. In this paper, we also propose a forecasting-based model for attack detection based on the collected multivariate time series data.

\subsection{Energy-based generative models}
Energy-based generative models (EBMs) are also called non-normalized probabilistic models, which directly estimate probability density with an unknown normalizing constant (\cite{song2021train}). Compared with other probabilistic methods like VAE (\cite{kingma2013auto}) and normalizing flows (\cite{rezende2015variational}), EBMs do not require tractability of the normalizing constants. Instead of specifying a normalized probability, they only estimate the unnormalized negative log-probability, which is called \textit{energy function}. Therefore, EBMs are more flexible and less restrictive in functional forms. 

Although EBMs have significant modeling advantages, exact likelihood is usually intractable and the inference process is usually slow. Recently some works \cite{san2021noise,lam2021bilateral} that aim to construct a new inference process have been proposed for acceleration. Besides, the unknown normalizing constant will make the training particularly difficult. There are currently three categories for training EBMs: (1) Noise contrastive estimation \cite{gutmann2010noise,gutmann2012bregman,bose2018adversarial} is used to learn EBMs by contrasting it with another known density. (2) Maximum likelihood training with MCMC. (3) Score matching-based methods (\cite{hyvarinen2005estimation}). Instead of estimating the log probability density functions (PDFs), they aim to estimate the first derivatives of the log-PDF. In this paper, we adopt score matching-based methods for training, which achieve state-of-the-art results (\cite{ho2020denoising,song2020score}) on image generation tasks recently.%TODO: 这里实际上并不是score matching,下面损失的等价求解推导 热力图： https://www.cxymm.net/article/qq_45866407/107891518

\subsection{Graph neural networks}
Graph neural networks (GNNs) have achieved impressive results on representation learning over non-Euclidean data (\cite{zhou2020graph}). They can well deal with the dependencies of different nodes. There are now two mainstreams of GNNs (\cite{wu2020comprehensive}): spectral-based methods and spatial-based methods. The main difference lies in the design of the kernel.  In this paper, we use graph attention networks (\cite{velivckovic2017graph}) to explicitly model the correlation between features of multivariate time series.
%CPS attack detection; anomaly deteciton; energy-based generative models; Graph neural networks.

\section{Preliminaries}\label{preliminaries}
In this section, attack detection architecture in CPSs is firstly introduced. Then graph attention network used for feature pattern extraction is presented. In the end, the denoising score matching method is presented. 
%TODO: energy based generative models
\subsection{Attack detection in CPSs}
\begin{figure}[t]
	\centering
	\includegraphics[width=0.6\textwidth]{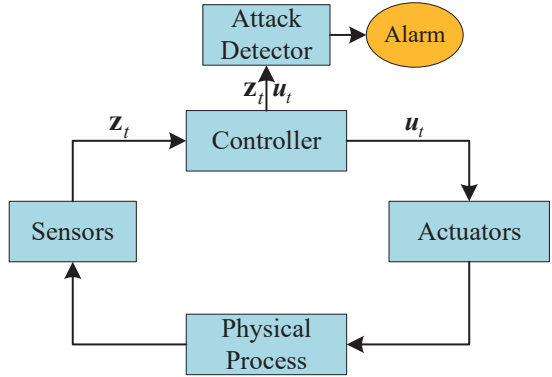}
	\caption{Anomaly detection architecture in CPSs.}
	\label{Fig. 1}
\end{figure}

CPSs are integrations of computation, networking, control and physical processes \cite{lee2017introduction}. As shown in Fig. \ref{Fig. 1}, sensors convert observations from the physical process to electronic signals. The controller sends control commands to actuators according to the received signals. The actuators convert control commands to mechanical motion. To monitor the status of CPSs, a commonly used anomaly detection architecture is shown in Fig. \ref{Fig. 1}. The control signals and sensor measurements are collected and monitored by an anomaly detector.

\subsection{Graph attention networks}
Consider a graph $\mathcal{G}=\{\mathcal{V}, \mathcal{H}\}$, where $\mathcal{V}$ and $\mathcal{H}$ are node vector and corresponding feature matrix respectively. GAT injects the graph structure by performing \textit{masked attention}. Assume $\mathcal{N}_i$ is the neighborhood of node $v_i$, the attention coefficients can be calculated as 
\begin{equation}
\begin{aligned}
\textbf{e}_{ij} &= \text{LeakyRELU} (\textbf{W}(\textbf{h}_i\oplus \textbf{h}_j))\\
\boldsymbol{\alpha}_{ij} &= \frac{\text{exp}(\textbf{e}_{ij})}{\sum_{k\in \mathcal{N}_i}\text{exp}(\textbf{e}_{ik})}\\
\end{aligned}
\end{equation}
where $\textbf{W}$ is a trainable matrix, $\textbf{h}_i$ is the feature vector of node $i$. $\oplus$ represents concatenation operation and  $\text{LeakyRELU}$ is 
\begin{equation}
\text{LeakyRELU}(x_i) =
\left\{
\begin{array}{lr}
x_i & \text{if}~ x_i > 0\\
\frac{x_i}{a_i}&\text{if}~ x_i < 0\\
\end{array}
\right., a_i >1
\end{equation}

In the end, the output of a GAT layer is 
\begin{equation}
o_i = \sigma(\sum_{j\in\mathcal{N}_i} \textbf{e}_{ij})
\end{equation}
where $\sigma$ represents sigmoid function.

\subsection{Score matching method}
\cite{hyvarinen2005estimation} proposes \textit{score matching} for learning non-normalized statistical models. Instead of using maximum likelihood estimation, they minimize the distance of derivatives of the log density function between data and estimated distributions. Denote $\nabla_x\log p_{\theta}(x)$ as \textit{score function} $s_{\theta}(x)$, then with a simple trick of partial integration, the objective function can be simplified as 
\begin{equation}\label{sm}
	\begin{aligned}
		J(\theta) &= \frac{1}{2}\mathbb{E}_{p(x)}\Vert \nabla_x\log p_{\theta}(x) - \nabla \log p(x)\Vert_2^2\\
		&= \mathbb{E}_{p(x)}\big[\text{Tr}(\nabla_x^2\log p_{\theta}(x)) + \frac{1}{2}\Vert\nabla_x\log p_{\theta}(x)\Vert_2^2\big] + \text{const.}\\
		&= \mathbb{E}_{p(x)}\big[\text{Tr}(\nabla_x s_{\theta}(x)) + \frac{1}{2}\Vert s_{\theta}(x)\Vert_2^2\big] + \text{const.}\\
	\end{aligned}
\end{equation}
where $p(x)$ and $p_{\theta}(x)$ are the data distribution and estimated distribution, respectively. $\theta$ represents trainable parameters. The estimated distribution will be equal to data distribution when Eq. \ref{sm} takes the minimum value.

However, Eq. \ref{sm} is difficult to calculate. \cite{vincent2011connection} connects denoising autoencoders and score matching and proposes \textit{denoising score matching} for density estimation. It firstly adds a bit of noise to each data: $\tilde x= x+\epsilon$, then the objective can be formulated as 
\begin{equation}\label{dsm}
	\mathbb{E}_{p(\tilde x, x)}\big[\frac{1}{2}\Vert\nabla_x\log p(\tilde x|x) - \nabla_x\log p_{\theta}(\tilde x)\Vert_2^2\big] + \text{const.}
\end{equation}

The expectation is approximate by the average of samples and the second derivatives do not need to be calculated compared with Eq. \ref{sm}.

\section{Methodology}\label{methods}
In this section, symbols and the problem definition will be firstly introduced. Then the basic procedure for anomaly detection and the framework of TFDPM are presented. In the end, an extra model is trained for accelerating prediction.

\subsection{Symbols and problem definition}
Consider a multivariate time series $\mathcal{X}=\{\boldsymbol{x}_1^0, \boldsymbol{x}_2^0, \cdots, \boldsymbol{x}_T^0\}\in\mathbb{R}^{T\times D}$, where $T$ is the total length and $D$ is the number of features. $\boldsymbol{x}_t^0$ represents the collected data from a CPS, which is $\boldsymbol{x}_t=\{\boldsymbol{z}_t, \boldsymbol{u}_t\}$. Our target is to determine whether the system is attacked given historical data up to time $t-1$ and observation at time $t$. Prediction-based methods are used to learn the normal pattern of data in the CPS. Specifically, sliding window data $\{\boldsymbol{x}_{t-\omega}^0, \boldsymbol{x}_{t-\omega+1}^0, \cdots, \boldsymbol{x}_{t-1}^0\}$ is used to predict $\boldsymbol{x}_t^0$ and the prediction method is shown in Eq. \ref{16}. Then attacks are detected based on the difference between predicted values and actual observations.
\begin{equation} \label{16}
q_{\mathcal{X}}(\boldsymbol{x}_{t_0:T}^0|\boldsymbol{x}_{t_0-\omega:t_0-1}^0) = \prod_{t=t_0}^T q_{\mathcal{X}}(\boldsymbol{x}_{t}^0|\boldsymbol{x}_{t-\omega:t-1}^0)
\end{equation}

\subsection{Overall structure}

\begin{figure}[t]
	\centering
	\includegraphics[width=0.9\textwidth]{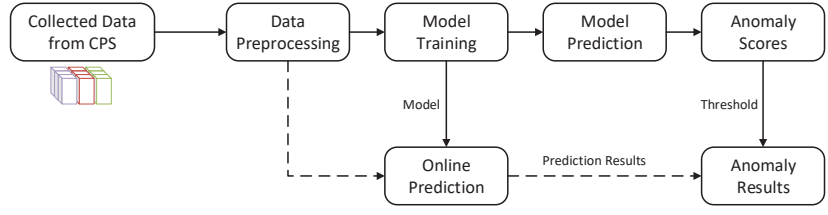}
	\caption{Overall structure of TFDPM.}
	\label{Fig. 2}
\end{figure}

As shown in Fig. \ref{Fig. 2}, TFDPM consists of three modules: data preprocessing, offline training and online detection. The data preprocessing module is shared by offline training and online detection. In this module, we discard missing values and encode discrete signals with one-hot vectors. Data is processed with min-max normalization firstly as shown in Eq. \ref{norm}. Then it's segmented into subsequences with sliding windows of length $\omega$. The processed data is sent to the training and prediction module. In the end, the difference between predicted values and actual observations are served as anomaly scores. A proper threshold can be selected for attack detection. In the end, the trained model and selected threshold are deployed for online detection.

\begin{equation}\label{norm}
\hat{x}_t^0=\frac{x_t^0 - \min X}{\max X - \min X}
\end{equation}

\subsection{Network architecture}
\begin{figure}[t]
	\centering
	\includegraphics[width=0.8\textwidth]{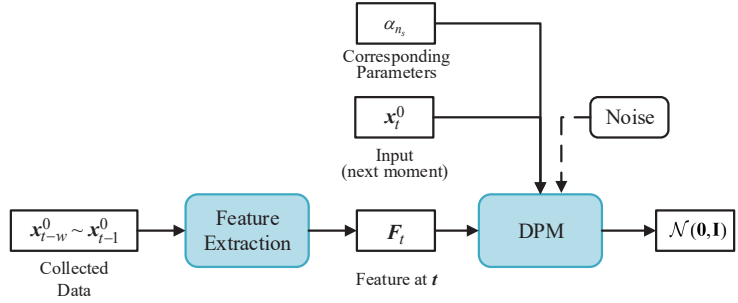}
	\caption{Network architecture of TFDPM at time step $t$. Feature at $t$ is obtained through a feature extraction module based on historical data. Then the input $\boldsymbol{x}_t^0$, feature $\boldsymbol{F}_t^0$ and corresponding parameters at step $n_s$ are sent to the conditional diffusion probability models for training.}
	\label{Fig. 3}
\end{figure}

The network architecture of TFDPM at time step $t$ is shown in Fig. \ref{Fig. 3}. It consists of two modules at each time step: temporal and feature pattern extraction module and conditional diffusion probabilistic generative module.
\subsubsection{Temporal pattern and feature pattern extraction} 
\begin{figure}[t]
	\centering
	\includegraphics[width=0.98\textwidth]{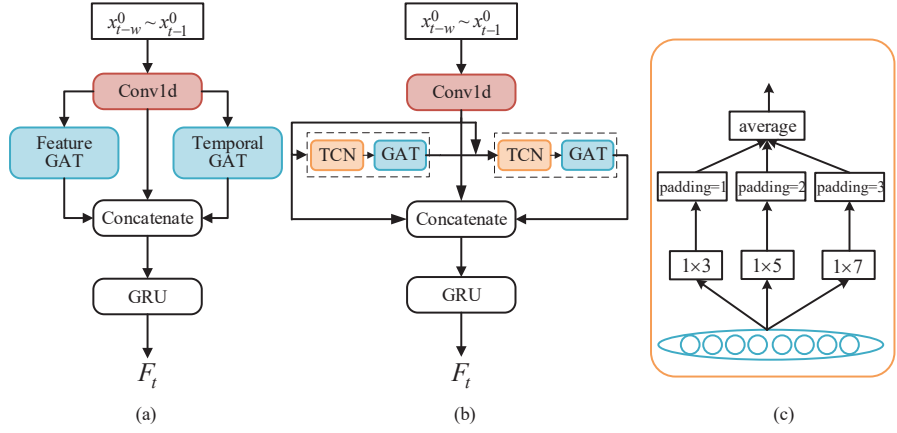}
	\caption{Temporal and feature pattern extraction methods. (a) \textbf{Double-GAT}: Use time oriented GAT and feature-oriented GAT to obtain representations firstly, then GRU is used for further feature extraction. (b) \textbf{TCN-GAT}: Use TCN and feature-oriented GAT to obtain representations firstly, then GRU is used for further feature extraction. (c) The architecture of TCN used in TCN-GAT. It adopts filter sizes of $1\times 3$, $1\times 5$, $1\times 7$ for convolution.}
	\label{Fig. 4}
\end{figure}

As shown in Fig. \ref{Fig. 3}, the module extracts feature $F_t$ given historical data up to time $t-1$. It's a general module and many methods can be used for feature extraction. In this paper, in order to explicitly model the relationship between different channels, we use graph attention networks (GAT) to extract features. Graph attention networks can also be used for temporal pattern extraction. In addition, temporal convolutional networks (TCN) \cite{oord2016wavenet} with dilations can also be used to capture temporal patterns. %is used to extract temporal features based on the historical data. 
As shown in Fig. \ref{Fig. 4},  we construct the following two networks for temporal pattern and feature pattern extraction module.

\textbf{Double-GAT} uses one-dimensional convolution with kernel size 5 to smooth the data. Then two GATs are used to obtain temporal and feature-oriented representations. Then we concatenate the output of GATs and the output of one-dimensional convolution and send them to GRU for feature extraction.

One-dimensional convolution with kernel size 5 is also carried out firstly to smooth the data in \textbf{TCN-GAT}. Different from Double-GAT, temporal convolutional layers (TCN) are used for temporal pattern extraction. As shown in Fig. \ref{Fig. 4} (c), in order to capture temporal features from different scales, we use three convolutional layers with filter sizes of $1\times 3$, $1\times 5$, $1\times 7$. In addition, different paddings are used to align the outputs of the convolutional layers. The output of TCN is the averaged value of the outputs of convolutional layers. 

As shown in Fig. \ref{Fig. 4} (b), the module that consists of a TCN and graph attention layer forms a single block of TCN-GAT. Two blocks are used here for capturing temporal and feature representations. The input of the second block is the average value of the output of the first block and the output of the 1-D convolutional layer. Similarly, we concatenate the outputs of the blocks and the 1-D convolution and send them to GRU for further feature extraction. 

\subsubsection{Conditional diffusion probabilistic model}
As shown in Fig. \ref{Fig. 3}, the extracted feature at each time step serves as the condition for the conditional diffusion probabilistic model. Then the prediction problem in Eq. \ref{16} can be approximated by Eq. \ref{17}, where $\theta$ represents trainable parameters of the temporal and feature pattern extraction module and the conditional diffusion probabilistic model.

\begin{equation}\label{17}
\prod_{t=t_0}^T q_{\mathcal{X}}(\boldsymbol{x}_{t}^0|\boldsymbol{x}_{t-\omega:t-1}^0)=\prod_{t=t_0}^Tp_{\theta}(\boldsymbol{x}_t^0|\boldsymbol{F}_{t})
\end{equation}

Let $\boldsymbol{x}_t^0\sim q_{\mathcal{X}}(\boldsymbol{x}_t^0|\boldsymbol{F}_t)$ denote the distribution of the dataset given $\boldsymbol{F}_t$, $p_{\theta}(\boldsymbol{x}_t^0|\boldsymbol{F}_t)$ denote a parameterized function that used to approximate $q_{\mathcal{X}}(\boldsymbol{x}_t^0|\boldsymbol{F}_t)$. Inspired by diffusion probabilistic models \cite{ho2020denoising}, we take a sequence of latent variables $\boldsymbol{x}_t^{1:N_s}$ to estimate $p_{\theta}(\boldsymbol{x}_t^0|\boldsymbol{F}_t)$, which is $p_{\theta}(\boldsymbol{x}_t^0|\boldsymbol{F}_t)=\int p_{\theta}(\boldsymbol{x}_t^{0:N_s}|\boldsymbol{F}_t)d\boldsymbol{x}_t^{1:N_s}$. If the approximated posterior satisfies Markov property, then it can be factorized as 
\begin{equation}
q(\boldsymbol{x}_t^{1:N_s}|\boldsymbol{x}_t^{0})=\prod_{n=1}^{N_s}q(\boldsymbol{x}_t^{n}|\boldsymbol{x}_t^{n-1}).
\end{equation}

The Markov chain,  which gradually adds Gaussian noise to the input, is constructed based on a variance schedule $\beta_{1:N_s}\in(0, 1)$. The transition operator is formulated as 
\begin{equation}
q(\boldsymbol{x}_t^{n}|\boldsymbol{x}_t^{n-1}) = \mathcal{N}(\boldsymbol{x}_t^n; \sqrt{1-\beta_{n}}\boldsymbol{x}_t^{n-1}, \beta_{n}\textbf{I})
\end{equation}
where $\sqrt{1-\beta_{n}}\boldsymbol{x}^{n-1}_t$ and $ \beta_{n}I$ are the expectation and variance of the Gaussian distribution. Denote $\alpha_n=1 - \beta_{n}$, $\bar{\alpha}_{n}=\prod_{i=1}^n\alpha_i$, then the conditional distribution of $\boldsymbol{x}_t^n$ given $\boldsymbol{x}_t^0$ can be obtained as 
\begin{equation}\label{6}
q(\boldsymbol{x}_t^{n}|\boldsymbol{x}_t^{0}) = \mathcal{N}(\boldsymbol{x}_t^n;\sqrt{\bar{\alpha}_n}\boldsymbol{x}_t^0, (1-\bar{\alpha}_n)\textbf{I}).
\end{equation}

%What's more, the 

With enough steps of state transition, the distribution of the last state gradually tends to standard Normal distribution, which is $p(\boldsymbol{x}_t^{N_s})\sim \mathcal{N}(\textit{0}, \textbf{I})$. If the \textit{reverse process} $p_{\theta}(\boldsymbol{x}_t^{0:N_s}|\boldsymbol{F}_t)$ is constructed similarly with that of the approximated posterior%, that is 
\begin{equation}
p_{\theta}(\boldsymbol{x}_t^{0:N_s}|\boldsymbol{F}_t) = p(\boldsymbol{x}_t^{N_s})\prod_{n=N_s}^{n=1}p_{\theta}(\boldsymbol{x}_t^{n-1}|\boldsymbol{x}_t^n, \boldsymbol{F}_t),
\end{equation}
\begin{equation}
p_{\theta}(\boldsymbol{x}_t^{n-1}|\boldsymbol{x}_t^n,\boldsymbol{F}_t) = \mathcal{N}(\boldsymbol{x}_t^n|\boldsymbol{\mu}_{\theta}(\boldsymbol{x}_t^n, \alpha_n,\boldsymbol{F}_t), \Sigma_{\theta}(\boldsymbol{x}_t^n, \alpha_{n}, \boldsymbol{F}_t)\textbf{I}),
\end{equation}
then according to Jensen's inequality, the conditional log-likelihood function can be estimated by 
\begin{equation}\label{elbo}
\begin{aligned}
\mathbb{E}\big(\log p(\boldsymbol{x}_t^{0}|\boldsymbol{F}_t)\big) &\geq \mathbb{E}_{q}\big(-\log p_{\theta}(\boldsymbol{x}_t^{0:N_s}|\boldsymbol{F}_t) + \log q(\boldsymbol{x}^{1:N_s}|\boldsymbol{x}_0, \boldsymbol{F}_t)\big)\\
&=\mathbb{E}_{q} \big[D_{\text{KL}}(q(\boldsymbol{x}_t^{N_s}|\boldsymbol{x}_t^0)\Vert p(\boldsymbol{x}_t^{N_s})) - \log p_{\theta}(\boldsymbol{x}_t^0|\boldsymbol{x}_t^1,\boldsymbol{F}_t)\\
&~~+ \sum_{n>1}D_{\text{KL}}(q(\boldsymbol{x}_t^{n-1}|\boldsymbol{x}_t^{n}, \boldsymbol{x}_t^0)\Vert p_{\theta}(\boldsymbol{x}_t^{n-1}|\boldsymbol{x}_t^n, \boldsymbol{F}_t)) \big].
\end{aligned}
\end{equation}

The first term can be viewed as a constant. In addition, the second term can be parameterized as $\boldsymbol{\mu}_{\theta}(\boldsymbol{x}_t^0, \alpha_1, \boldsymbol{F}_t)$. Therefore, we only need to calculate $D_{KL}(q(\boldsymbol{x}^{n-1}|\boldsymbol{x}^{n}, \boldsymbol{x}^0)\Vert p_{\theta}(\boldsymbol{x}_t^{n-1}|\boldsymbol{x}_t^n)$. Fortunately, $q(\boldsymbol{x}_t^{n-1}|\boldsymbol{x}_t^{n}, \boldsymbol{x}_t^0)$ in the third term is tractable due to the good property of Gaussian distribution, that is 
\begin{equation}
q(\boldsymbol{x}_t^{n-1}|\boldsymbol{x}_t^{n}, \boldsymbol{x}_t^0) = \mathcal{N}(\boldsymbol{x}_t^{n-1}; \tilde{\boldsymbol{\mu}}_n(\boldsymbol{x}_t^n, \boldsymbol{x}_t^0), \tilde{\beta}_n\textbf{I}),
\end{equation}
where
\begin{equation}\label{11}
\tilde{\boldsymbol{\mu}}_n(\boldsymbol{x}_t^n, \boldsymbol{x}_t^0) = \frac{\sqrt{\bar{\alpha}_{n-1}}\beta_n}{1-\bar{\alpha}_n}\boldsymbol{x}_t^0 + \frac{\sqrt{\alpha_n}(1-\bar{\alpha}_{n-1})}{1-\bar{\alpha}_n}\boldsymbol{x}_t^n,
\end{equation}
\begin{equation}\label{t_beta}
\tilde{\beta}_n = \frac{1-\bar{\alpha}_{n-1}}{1-\bar{\alpha}_n}\beta_n.
\end{equation}

%As shown in \cite{ho2020denoising}, 
If we set $\Sigma_{\theta}=\tilde{\beta}_n \textbf{I}$, the third term in Eq. \ref{elbo} can be simplified as 
\begin{equation}\label{13}
\begin{aligned}
D_{\text{KL}} & (q(\boldsymbol{x}_t^{n-1}|\boldsymbol{x}_t^{n}, \boldsymbol{x}_t^0)\Vert p_{\theta}(\boldsymbol{x}_t^{n-1}|\boldsymbol{x}_t^n, \boldsymbol{F}_t))\\ 
& =\mathbb{E}_q\big[\frac{1}{2\Sigma_{\theta}} \Vert\tilde{\boldsymbol{\mu}}_n(\boldsymbol{x}_t^n, \boldsymbol{x}_t^0) - \tilde{\boldsymbol{\mu}}_{\theta}(\boldsymbol{x}_t^n, \alpha_n, \boldsymbol{F}_t)\Vert^2\big] + C,
\end{aligned}
\end{equation}
where $C$ is a constant. What's more, combining Eq. \ref{6} and Eq. \ref{11}, one can get $\tilde{\boldsymbol{\mu}}_n(\boldsymbol{x}_t^n, \boldsymbol{x}_t^0) = (\boldsymbol{x}_t^n - \beta_n\boldsymbol{\epsilon}/\sqrt{1-\bar{\alpha}_n})/\sqrt{\alpha_n}$, where $\boldsymbol{\epsilon}\in \mathcal{N}(\textbf{0}, \textbf{I})$. If $\tilde{\boldsymbol{\mu}}_{\theta}$ is parameterized similar with $\tilde{\boldsymbol{\mu}}_n(\boldsymbol{x}^n, \boldsymbol{x}^0)$, that is $\tilde{\boldsymbol{\mu}}_{\theta}=(\boldsymbol{x}_t^n - \beta_n\boldsymbol{\epsilon}_{\theta}(\boldsymbol{x}_t^n, \alpha^n, \boldsymbol{F}_t)/\sqrt{1-\bar{\alpha}_n})/\sqrt{\alpha_n}$, then Eq. \ref{13} can be simplified as 
\begin{equation}
\mathbb{E}_{\boldsymbol{x}^{0}, \boldsymbol{\epsilon}}\big[\frac{\beta_n^2}{2\Sigma_{\theta}\alpha_n(1-\bar{\alpha}_n)}\Vert\boldsymbol{\epsilon} - \boldsymbol{\epsilon}_{\theta}(\boldsymbol{x}_t^n, \alpha_n, \boldsymbol{F}_t)\Vert^2\big].
\end{equation}
where $\boldsymbol{x}_t^n$ is parameterized by $\sqrt{\bar{\alpha}_n}\boldsymbol{x}_t^0 + \sqrt{1-\bar{\alpha}_n} \boldsymbol{\epsilon}$, $\boldsymbol{\epsilon}\sim \mathcal{N}(\textbf{0}, \textbf{I})$. In order to balance the noise and the signal,  we use a weighted variational lower bound $\mathcal{L}_t$ as training objective:
\begin{equation}\label{loss}
\mathcal{L}_t = \mathbb{E}_{\boldsymbol{x}_t^0, \boldsymbol{\epsilon}, n}\big[\frac{N_s}{2}(\text{SNR}(n-1)-\text{SNR}(n))\Vert\boldsymbol{\epsilon} - \boldsymbol{\epsilon}_{\theta}(\boldsymbol{x}_t^n, \alpha_n, \boldsymbol{F}_t)\Vert^2\big].
\end{equation}
where $\text{SNR}$ represents the \textit{signal-to-noise ratio} function and can be calculated as 
\begin{equation}\label{snr}
\text{SNR}(n) = \frac{\mathbb{E}^2(q(\boldsymbol{x}_t^n|\boldsymbol{x}_t^0))}{\text{Var}(q(\boldsymbol{x}_t^n|\boldsymbol{x}_t^0))} = \frac{\bar{\alpha}_n}{1-\bar{\alpha}_n}.
\end{equation}
%In addition, the connection between Eq. \ref{snr} and denoising score matching method in Eq. \ref{dsm} can be found in Appendix.
Actually, Eq. \ref{loss} can be viewed as a weighted version of the denoising score matching objective in Eq. \ref{dsm}, detail derivation can be found in \ref{A}.
% and directly optimizes the mean square error between $\boldsymbol{\epsilon}$ and $\boldsymbol{\epsilon}_{\theta}$, that is
%\begin{equation}
%\mathbb{E}_{\boldsymbol{x}_t^{0}, \boldsymbol{\epsilon}, n}\big[\Vert\boldsymbol{\epsilon} - \boldsymbol{\epsilon}_{\theta}(\boldsymbol{x}_t^n, \alpha_n, \boldsymbol{F}_t)\Vert^2\big].
%\end{equation}

For the trained conditional diffusion probabilistic model, the generation process depends on $p_{\theta}(\boldsymbol{x}^{n-1}|\boldsymbol{x}^n, \boldsymbol{F}_t)$, which can be calculated as 
\begin{equation}\label{15}
\boldsymbol{x}_t^{n-1} = \frac{1}{\sqrt{\alpha_{n}}}\big(\boldsymbol{x}_t^n - \frac{\beta_n}{\sqrt{1-\bar{\alpha}_n}}\boldsymbol{\epsilon}_{\theta}(\boldsymbol{x}_t^n, \alpha_n, \boldsymbol{F}_t) + \sqrt{\Sigma_{\theta}} \boldsymbol{z}\big)
\end{equation}
where $\boldsymbol{z}\sim \mathcal{N}(\textbf{0}, \textbf{I})$. All in all, starting from a sample from white noise $\boldsymbol{x}^N$, $\boldsymbol{x}^0$ can be reconstructed by iteratively calling Eq. \ref{15} $N_s$ times.

%The conditional evidence lower bound at each time step $\mathcal{L}_t$ can be obtained through a similar derivation as in section \ref{dpm}, the $\boldsymbol{\epsilon}_{\theta}$ is also conditioned on the feature state at time $t$.
%\begin{equation}
%\mathcal{L}_t = \mathbb{E}_{\boldsymbol{x}_t^0, \boldsymbol{\epsilon}, n}\big[\Vert\boldsymbol{\epsilon} - \boldsymbol{\epsilon}_{\theta}(\boldsymbol{x}_t^n, \boldsymbol{F}_t, \alpha_n)\Vert^2\big].
%\end{equation}
%
%Inspired by \cite{kingma2021variational}, we introduce an weighted version of $\mathcal{L}_t$, 

\begin{figure}[t]
	\centering
	\includegraphics[width=0.98\textwidth]{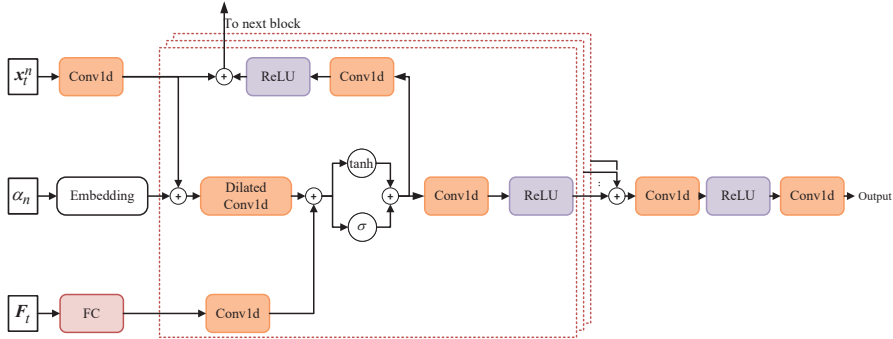}
	\caption{Network architecture of the conditional $\boldsymbol{\epsilon}_{\theta}$ network.}
	\label{Fig. 5}
\end{figure}

As for the $\boldsymbol{\epsilon}_{\theta}$ network, we design a network inspired by WaveNet \cite{oord2016wavenet}. The network consists of 4 residual blocks and the architecture of a single block is shown in Fig. \ref{Fig. 5}. We firstly transform $\alpha_n$ to $[0, 1]$ and use Fourier feature embeddings (\cite{tancik2020fourier}) for $\alpha_n$. The observed values are smoothed with 1-dimensional convolution and we sum the output with the embeddings and send them to a 1-dimensional dilated convolutional layer firstly. Secondly, the sum of the output and the feature extracted from the condition $\boldsymbol{F}_t$ are served as input for the gated activation unit. In the end, one part of the output is fed into a 1-dimensional convolutional layer and served as the output of the block while the other part is summed with the feature extracted from observations and served as the input of the next block.

\subsection{Training}
The training procedure at each time step $t$ is shown in Algorithm \ref{train}. Given the historical data $\boldsymbol{x}_{t-\omega:t-1}^0$, the feature state $\boldsymbol{F}_t$ is first extracted based on temporal and feature pattern extraction module. Then we get samples via the tractable distribution $q(\boldsymbol{x}_t^n|\boldsymbol{x}_{t}^0)$. In addition, the \textit{signal-to-noise ratio} at certain step $n$ can be obtained given $\alpha_{1:N_s}$. In the end, the loss function $\tilde{\mathcal{L}}_t(\theta)$ can be calculated according to Eq. \ref{loss}, where $\theta$ represents the trainable parameters in TFDPM.

\begin{algorithm}[tb]
	\caption{Training procedure at time step $t$.}
	\label{train}
	\textbf{Input}: Sliding window data $\boldsymbol{x}_{t-\omega:t-1}^0$, observed data $\boldsymbol{x}_t^0$, noise scales $\alpha_{1:N_s}$
	\begin{algorithmic}[1]
		\STATE Get $\boldsymbol{F}_{t}$ based on temporal and feature pattern extraction module. $\boldsymbol{F}_t = \text{TCN-GAT}(\boldsymbol{x}_{t-\omega:t-1}^0)$ or $\boldsymbol{F}_t = \text{Double-GAT}(\boldsymbol{x}_{t-\omega:t-1}^0)$ 
		\WHILE {not converged}
		\STATE Initialize $n\sim \text{Discrete Uniform}(1, N_s)$, $\boldsymbol{\epsilon}\sim\mathcal{N}(\textbf{0},\textbf{I})$
		\STATE Calculate $\boldsymbol{x}_t^n$ via $\sqrt{\bar{\alpha}_n}\boldsymbol{x}_t^0 + \sqrt{1-\bar{\alpha}_n} \boldsymbol{\epsilon}$
		\STATE Calculate $\text{SNR}(n)$ and $\text{SNR}(n-1)$ according to Eq. \ref{snr}.
%		\STATE Get samples at $t_s$: $\bm{x}_t^{t_s} = \bm{m}_t^{t_s} + \bm{z}*\bm{v}_t^{t_s}$.
		\STATE Calculate loss $\mathcal{L}_t(\theta)$ according to Eq. \ref{loss}.	
		\STATE Take gradient descent step on $\nabla_{\theta}\mathcal{L}_t(\theta)$
		\ENDWHILE
	\end{algorithmic}
\end{algorithm}

\subsection{Prediction and attack detection}
\begin{figure}[t]
	\centering
	\includegraphics[width=0.8\textwidth]{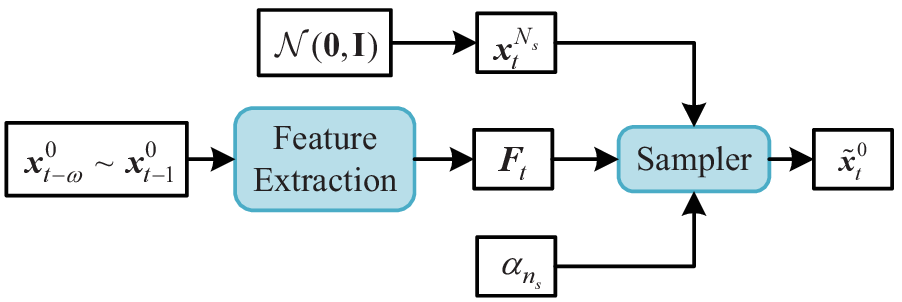}
	\caption{Prediction of $\boldsymbol{x}_t^0$ via historical data $\boldsymbol{x}_{t-\omega:t-1}^0$.}
	\label{Fig. 6}
\end{figure}

The prediction process is shown in Fig. \ref{Fig. 6}. Feature state $\boldsymbol{F}_t$ is firstly obtained based on historical data $\boldsymbol{x}_{t-\omega:t-1}^0$. $\boldsymbol{x}_t^{N_s}$ can be sampled from the target distribution $\mathcal{N}(\textbf{0}, \textbf{I})$. Then the feature state $\boldsymbol{F}_t$, samples from target distribution $\boldsymbol{x}_t^{N_s}$ and parameters at corresponding step $\alpha_{n_s}$ are sent to the sampler of diffusion probabilistic models for prediction. The detailed procedure of the sampler at $t$ is shown in Algorithm \ref{sampler}.

\begin{algorithm}[tb]
	\caption{Procedure of sampler at time step $t$}
	\label{sampler}
	\textbf{Input}: Feature state $\boldsymbol{F}_t$, $\alpha_{1:N_s}$, $\tilde{\beta}_{1:N_s}$
	\begin{algorithmic}[1]
		\STATE $\boldsymbol{x}_t^{N_s}\sim \mathcal{N}(\textbf{0}, \textbf{I})$
		\FOR {$n=N_s, \cdots, 1$}
		\STATE $ \boldsymbol{z}\sim \mathcal{N}(\textbf{0}, \textbf{I})$ if $n>1$ else $\boldsymbol{z}=\textbf{0}$
		\STATE $\boldsymbol{x}^{n-1}_t = \frac{1}{\sqrt{\alpha_{n}}}\big(\boldsymbol{x}^n_t - \frac{\beta_n}{\sqrt{1-\bar{\alpha}_n}}\boldsymbol{\epsilon}_{\theta}(\boldsymbol{x}_t^n, \alpha_n, \boldsymbol{F}_t) + \tilde{\beta}_n \boldsymbol{z}\big)$
		\ENDFOR
		\STATE return $\boldsymbol{x}_t^0$
	\end{algorithmic}
\end{algorithm}

Let $\tilde{\boldsymbol{x}}_{ti}^0$ denote the predicted value of the i-th feature at time $t$. In this paper, the mean square error (MSE) between the predicted value and the actual observation at each time step serves as \textit{anomaly scores}, which can be obtained via Eq. \ref{rmse}. A larger anomaly score indicates the system is more likely to be attacked at the corresponding time.  Denote the anomaly scores on the test set as $\{l_{1}, l_2, \cdots l_Q\}$, where $Q$ represents the number of observations in the test set. Then anomalies can be detected by setting a threshold, which can be tuned by various methods. For example, \cite{siffer2017anomaly} proposes peaks over threshold (POT) algorithm based on extreme value theory. \cite{hundman2018detecting} proposes a dynamic thresholding method. These methods are not the focus of this paper and we simply use the best metrics obtained from the grid-search method for comparison.
\begin{equation}\label{rmse}
\text{l}_t = \sum^D_{i=1}\frac{(\tilde{x}_{ti}^0 - x_{ti}^0)^2}{D}
\end{equation}

\subsection{Efficient sampling}
The major limitation of diffusion probabilistic models is that transforming data to target distribution takes many diffusion steps, which makes the generation process slower than other generative models like VAEs and GANs. As shown in Algorithm \ref{sampler}, the conditional $\epsilon_{\theta}$ network has to be called $N_s$ times for each prediction step. To solve the problem, we propose a conditional noise scheduling network for the scheduling of noise sequences $\beta_{1:N_s}$. % inspired by \cite{lam2021bilateral}

Assume the learned noise scales are $0<\hat\beta_{1}<\cdots<\hat\beta_{N_l}<1$, where $N_l\ll N_s$. Similarly, we can get  $\hat{\alpha}_n = 1-\hat\beta_n$, $\hat{\bar{\alpha}}_n=\prod_{i=1}^{n}\hat{\alpha}_i$. In addition, it's easy to obtain that $\hat\beta_n$ satisfies $0<\hat\beta_n < \min \left\{1-\frac{\hat{\bar{\alpha}}_{n+1}}{1-\hat\beta_{n+1}}, \hat\beta_{n+1}\right\}$. As a result, we can design a neural network $\sigma_{\phi}(\boldsymbol{x}_t^n,\boldsymbol{F}_t)\in(0, 1)$ such that
\begin{equation}\label{bt}
\hat\beta_n = \min \left\{1-\frac{\hat{\bar{\alpha}}_{n+1}}{1-\hat\beta_{n+1}}, \hat\beta_{n+1}\right\}\sigma_{\phi}(\boldsymbol{x}_t^n, \boldsymbol{F}_t)
\end{equation}
where $\phi$ represents trainable parameters in $\sigma_{\phi}$. In order to find a shorter noise schedule, we set $q_{\phi}(\boldsymbol{x}_{n+1}|\boldsymbol{x}_n)=q(\boldsymbol{x}_{n_1+\tau}|\boldsymbol{x}_{n_1})$, where $\tau \in(1,N_s)$ is a positive integer. The constraint makes one step of diffusion using $\hat{\beta}_n$ equal to $\tau$ steps of diffusion using $\beta_{n_1:n_1+\tau}$.

The training of the noise scheduling network depends on the trained TFDPM. We want to optimize the gap between the log-likelihood and the variational lower bound given the condition $\boldsymbol{F}_t$ at each time step. Firstly, for all $n\in\{2,\cdots, N_s\}$, the log-likelihood function can be estimated by a new variational lower bound $F_{\text{score}}^{t, n}(\theta)$ which is equivalent to the lower bound in Eq. \ref{elbo}.
\begin{equation}
\begin{aligned}
\mathbb{E}(\log p(\boldsymbol{x}_t^0|\boldsymbol{F}_t)) \geq &-\mathbb{E}_{q(\boldsymbol{x}_t^n|\boldsymbol{x}_t^0)}\big[D_{\text{KL}}(p_{\theta}(\boldsymbol{x}_t^{n-1}|\boldsymbol{x}_t^n, \boldsymbol{F}_t)\Vert q(\boldsymbol{x}_t^{n-1}|\boldsymbol{x}_t^n, \boldsymbol{x}_t^0))\\
&-\mathbb{E}_{p_{\theta}(\boldsymbol{x}_t^1|\boldsymbol{x}_t^n)}\log p_{\theta}(\boldsymbol{x}_t^0|\boldsymbol{x}_t^1, \boldsymbol{F}_t)\big]=F_{\text{score}}^{t, n}(\theta).
\end{aligned}
\end{equation}%TODO: appendix给出推导过程来

Denote the optimal parameters of TFDPM as $\theta^*$, we have $p_{\theta^*}(\boldsymbol{x}_{t}^{0:n-1}|\boldsymbol{F}_t) = q(\boldsymbol{x}_{t}^{0:n-1}|\boldsymbol{F}_t)$. When $\tau=1$, $q_{\phi}(\boldsymbol{x}_t^n|\boldsymbol{x}_t^{n-1}, \boldsymbol{F}_t)=q(\boldsymbol{x}_t^n|\boldsymbol{x}_t^{n-1}, \boldsymbol{F}_t)$, the gap between $\log p(\boldsymbol{x}_t^0|\boldsymbol{F}_t)$ and $F_{\text{score}}^{t, n}(\theta^*)$ can be formulated as 
\begin{equation}
\log p(\boldsymbol{x}_t^0|\boldsymbol{F}_t) - F_{\text{score}}^{t, n}(\theta^*) = \mathbb{E}_{q(\boldsymbol{x}_t^n|\boldsymbol{x}_t^0)}\big[\sum^n_{i=2}\mathcal{L}^{(i)}(\phi; \theta^*)\big]
\end{equation}
where $\mathcal{L}^{(i)}(\phi; \theta^*)=D_{\text{KL}}(p_{\theta^*}(\boldsymbol{x}_t^{i-1}|\boldsymbol{x}_t^i, \boldsymbol{F}_t)\Vert q_{\phi}(\boldsymbol{x}_t^{i-1}|\boldsymbol{x}_t^0, \boldsymbol{F}_t))$. Detail derivation can be found in \ref{B}. It's worth noting that compared with $q(\boldsymbol{x}_t^n|\boldsymbol{x}_t^0)$ in TFDPM, $\boldsymbol{F}_t$ is also used as the condition in $q_{\phi}(\boldsymbol{x}_t^{i-1}|\boldsymbol{x}_t^0, \boldsymbol{F}_t)$ when learning new noise scales because $\hat\beta_{n}$ depends on $\boldsymbol{F}_t$ and $\boldsymbol{x}_t^n$ simultaneously. The new lower bound can be formulated as
\begin{equation}
	\mathcal{L}_{\text{bdm}} = F_{\text{score}}^{t, n}(\theta^*) + \mathbb{E}_{q(\boldsymbol{x}_t^n|\boldsymbol{x}_t^0)}\big[\mathcal{L}^{(i)}(\phi; \theta^*)\big]
\end{equation}

Substitute $p_{\theta^*}(\boldsymbol{x}_t^{i-1}|\boldsymbol{x}_t^i, \boldsymbol{F}_t)$ and $q_{\phi}(\boldsymbol{x}_t^{i-1}|\boldsymbol{x}_t^0, \boldsymbol{F}_t)$ in $\mathcal{L}^{(i)}(\phi; \theta^*)$, it can be simplified as 
%It adopts weighted mean square error as training objective of the noise scheduling network, which is shown in 
Eq. \ref{ns}. The corresponding training procedure is shown in Algorithm \ref{train_ns}.  Since the model contains a bilateral modeling objective for both score network and noise scheduling network, we call this method as $\text{TFBDM}$.

\begin{equation}\label{ns}
\mathcal{L}^{(i)}(\phi;\theta^*)=\frac{1}{2(1-\hat\beta_n - \hat{\bar{\alpha}}_n)} \Vert \sqrt{1-\hat{\bar{\alpha}}_n}\boldsymbol{\epsilon}_n - \frac{\hat\beta_n}{\sqrt{1-\hat{\bar{\alpha}}_n}}\boldsymbol{\epsilon}_{\theta^*}(\boldsymbol{x}^n_t, \boldsymbol{F}_t, \hat{\alpha}_n)\Vert_2^2 + C_n(\phi)
\end{equation}
where 
\begin{equation}
C_n(\phi) = \frac{1}{4}\log \frac{1-\hat{\bar{\alpha}}_n}{\hat\beta_n} + \frac{D}{2}\big(\frac{\hat\beta_n}{1-\hat{\bar{\alpha}}_n} -1\big).
\end{equation}

\begin{algorithm}[tb]
	\caption{Training procedure of noise scheduling network.}
	\label{train_ns}
	\textbf{Input}: Given $\boldsymbol{\epsilon}_{\theta^*}$, $\tau$, $\beta_{1:N_s}$,  $\alpha_{1:N_s}$, $\boldsymbol{x}_{t-\omega:t-1}^0$
	\begin{algorithmic}[1]
		\WHILE {not converged}
		\STATE Get  $\boldsymbol{F}_t$ based on historical data $\boldsymbol{x}_{t-\omega:t-1}^0$.
		\STATE Initialize $n\sim \text{Discrete Uniform}(2, N_s-\tau)$, $\boldsymbol{\epsilon}\sim\mathcal{N}(\textbf{0},\textbf{I})$
		\STATE $\hat{\alpha}_n = \alpha_{n}$, $\hat{\beta}_{n+1}=1-(\bar{\alpha}_{n+\tau}/\bar{\alpha}_n)$
		\STATE Calculate $\boldsymbol{x}_t^n$ via $\sqrt{\hat{\bar\alpha}_n}\boldsymbol{x}_t^0 + \sqrt{1-\hat{\bar\alpha}_n} \boldsymbol{\epsilon}$
		\STATE $\boldsymbol{\epsilon}_{\theta^*} = \boldsymbol{\epsilon}_{\theta^*}(\boldsymbol{x}^n_t, \boldsymbol{F}_t, \hat{\alpha}_n)$
		\STATE Get $\hat \beta_n$ via Eq. \ref{bt}.
		\STATE Calculate loss $\mathcal{L}(\phi)$ according to Eq. \ref{ns}.	
		\STATE Take gradient step on $\nabla_{\phi}\mathcal{L}(\phi)$
		\ENDWHILE
	\end{algorithmic}
\end{algorithm}

Once the conditional noise scheduling network is trained, we can sample from the target distribution based on the new noise schedule. Since the generation process starts from the last noise scale, we need to set two hyperparameters $\hat{\bar\alpha}_N$ and $\hat{\beta}_N$ firstly. The smallest noise scale $\beta_1$ is used as a threshold to determine when to stop the sampling process. In the end, the sampler can be modified as Algorithm \ref{t_sampler}.

\begin{algorithm}[tb]
	\caption{Procedures of sampler at time step $t$}
	\label{t_sampler}
	\textbf{Input}: Feature state $\boldsymbol{F}_t$, $\alpha_{1:N_s}$, $\hat{\bar{\alpha}}_{N}$, $\hat{\beta}_{N}$
	\begin{algorithmic}[1]
		\STATE $\boldsymbol{x}_t^{N_s}\sim \mathcal{N}(\textbf{0}, \textbf{I})$, $\hat\beta_{ls}=[\hat{\beta}_{N}]$
		\FOR {$n=N_s, \cdots, 2$}
		\STATE $\boldsymbol{z}\sim \mathcal{N}(\textbf{0}, \textbf{I})$, $\hat{\alpha}_n=1-\hat{\beta}_n$, get $\hat{\tilde\beta}_n$ according to Eq. \ref{t_beta}.
		\STATE $\boldsymbol{x}^{n-1}_t = \frac{1}{\sqrt{\hat{\alpha}_{n}}}\big(\boldsymbol{x}^n_t - \frac{\hat{\beta}_n}{\sqrt{1-\hat{\bar{\alpha}}_n}}\boldsymbol{\epsilon}_{\theta}(\boldsymbol{x}_t^n, \hat{\alpha}_n,\boldsymbol{F}_t) + \hat{\tilde{\beta}}_n \boldsymbol{z}\big)$
%		\STATE 
		\STATE $\hat{\bar\alpha}_{n-1} = \frac{\hat{\bar\alpha}_{n}}{\hat{\alpha}_n}$, $\hat\beta_{n-1} = \min \left\{1-\frac{\hat{\bar{\alpha}}_{n}}{1-\hat\beta_{n}}, \hat\beta_{n}\right\}\sigma_{\phi}(\boldsymbol{x}_t^{n-1})$
		\IF{$\hat{\beta}_{n-1} < \beta_1$}
		\STATE \textbf{break}
		\ENDIF
		\STATE Push $\hat{\beta}_{n-1}$ into the stack $\hat\beta_{ls}$.
		\ENDFOR
		\STATE Calculate parameters based on the constructed noise schedule $\hat\beta_{ls}$.
		\STATE Get samples $\boldsymbol{x}_t^0$ based on Algorithm \ref{sampler}.
		\STATE return $\boldsymbol{x}_t^{0}$.
	\end{algorithmic}
\end{algorithm}

\section{Experiments}\label{experiments}
In this section, TFDPM is evaluated on three real-world CPS datasets and we compare the results with some other state-of-the-art attack detection methods. Then we compare the performance of attack detection and prediction speed of TFBDM and TFDPM. In addition, we perform extensive experiments to show the effectiveness of our models.

\subsection{Datasets and evaluation metrics}
In this paper, we use the following three real-world CPS datasets for evaluation:
\begin{enumerate}
	\item \textbf{PUMP}: It consists of the data collected from a water pump system from a small town. The dataset is collected every minute for 5 months.
	\item \textbf{SWAT}: The dataset comes from a water treatment test-bed which is a small-scale version of the modern CPS \cite{goh2016dataset}. The system has been widely used and it's very important to detect potential attacks from malicious attackers. In our experiments, the original data samples are downsampled to a data point every 10 seconds.
	\item \textbf{WADI}: Water Distribution (WADI) is a distribution system which consists of many water  distribution pipelines \cite{ahmed2017wadi}. It forms a complex system that contains water treatment, storage and distribution networks. Two weeks of normal operations are used as training data. Similarly, we downsample the the samples to a data point every 10 seconds. In addition, the data in the last day is ignored because they have different distributions compared with the data in previous days.
\end{enumerate}

Detailed properties of these datasets are shown in Table \ref{table 1}. It shows the number of sensors and actuators in each dataset. Note that the labels of anomalies are only available in test sets. 

As for evaluation metrics, we use precision, recall and F1-score to compare the performance of TFDPM and other methods: $\text{F}1=\frac{2\times \text{precision}\times\text{recall}}{\text{precision}+\text{recall}}$, where $\text{precision}=\frac{\text{TP}}{\text{TP}+\text{FP}}$, $\text{recall}=\frac{\text{TP}}{\text{TP}+\text{FN}}$, $\text{TP}$,$\text{TN}$, $\text{FP}$ and $\text{FN}$ represent the number of true positives, true negatives, false positives and false negatives. What's more, the anomalous points are usually continuous and form many continuous anomaly segments \cite{su2019robust}. Therefore, the attacks are considered as successfully detected if an alert is sent within the continuous anomalous points segment. In this paper, we mainly compare the best F1-scores on various datasets which indicate the upper bound of the model capability.

\begin{table}[t]%TODO:检查这里的数据是否正确
	\centering
	\caption{Detail information of the CPS datasets.}
	\begin{tabular}{cccccc}
		\hline
		&\makecell[c]{TRAIN\\SIZE}&\makecell[c]{TEST\\ SIZE}&\makecell[c]{NUM. \\SENSORS}&\makecell[c]{NUM.\\ ACTUATORS}&ANOMALIES(\%)\\
		\hline
		\hline
		PUMP&76901&143401&44&0&10.05\\
		SWAT&49668&44981&25&26&11.97\\
		WADI&120960&15701&67&26&7.09\\
		\hline
	\end{tabular}
	\label{table 1}
\end{table}

\subsection{Competitive methods}
We compare the performance of our proposed method with the following state-of-the-art attack detection methods.
\begin{enumerate}
	\item \textbf{Isolation forest }: Isolation forest \cite{liu2008isolation} is an unsupervised attack detection method based on decision tree algorithms and has been successfully used in various fields. It usually has fast training speed and good generalization performance.
	\item \textbf{LSTM-AE}: LSTM-AE \cite{malhotra2016lstm} uses a LSTM-based autoencoder to model time series and detects attacks based on the difference between actual observations and reconstruction values.
	\item \textbf{Sparse-AE}: Sparse-AE \cite{ng2011sparse} also adopts an autoencoder for attack detection. Compared with LSTM-AE, it uses sparse hidden embeddings to get better latent representations of input data.
	\item \textbf{LSTM-PRED}: LSTM-PRED \cite{goh2017anomaly} uses LSTM-based methods for prediction. It detects attacks based on the difference between predicted values and actual observations.
	\item \textbf{DAGMM}: DAGMM \cite{zong2018deep} combines deep generative models and Gaussian mixture models, which can generate a low-dimensional representation for each observation. The attacks are also detected based on the reconstruction error.
	\item \textbf{OmniAnomaly}: OmniAnomaly \cite{su2019robust} is a stochastic recurrent neural network based on variational autoencoder. It models the normal pattern of time series and detects attacks based on reconstruction probability \cite{an2015variational}.
	\item \textbf{USAD}: USAD \cite{audibert2020usad} is a recently proposed method based on autoencoders. The architecture allows it to isolate anomalies and provides fast training. The adversarial training trick is also demonstrated helpful for attack detection.
	
\end{enumerate}

Since Sparse-AE and Isolation forest are not directly designed for time series anomaly detection, the observed values and states and actuators in $\omega$ time points are stacked as the inputs for the models.
\subsection{Hyperparameter settings}
% TS feature extraction, DDPM module, accelerating diffusion module
The length of the sequence of noise scales $N_s$, which is also called diffusion steps, is set as $100$ for all experiments. In addition, the sequence of noise scales for the forward process is set constants increasing from $\beta_1=10^{-4}$ to $\beta_{N_s}=10^{-2}$. In addition, the batch size for training is set as 100 for all three datasets.  The default length of sliding window $\omega$ is set as 12. TFDPM is trained for 20 epochs with early stopping.

As for the noise scheduling network, there are three important parameters, $\tau$, $\hat{\bar{\alpha}}_N$, $\hat{\beta}_{N}$. A larger $\tau$ makes the length of constructed noise scales shorter. We set $\tau=10$ for all datasets. In order to find a good set of initial values for $(\hat{\bar{\alpha}}_N, \hat{\beta}_{N})$, we set the range of these parameters from $0.1$ to $0.9$ and apply a grid search algorithm for parameter tuning. The noise scheduling network is also trained for 20 epochs with early stopping.

% TODO: 实验设置，实验结果比较，加速结果比较，超参数影响：batch-size， window-size, diffusion steps

\subsection{Results and Analysis}
In this section, we demonstrate that TFDPM outperforms other baselines based on the performance on three CPS datasets. Then the prediction speed and performance of TFBDM and TFDPM are compared. What's more, the influence of hyperparameters on the performance is explored.
% TODO: (1)异常检测结果 (2) 预测效果展示，最好有典型异常 (3) TFBDM效果展示 (4) 超参数的影响，batch_size, window_length, diffusion step
% TODO: 结果的阐述，
% ML-based methods与深度学习方法的比较，deterministic methods与概率生成模型效果的比较，我们的模型的比较
% ablation studies，与其他方法的比较
\subsubsection{Comparison of performance}
We use different temporal pattern and feature pattern extraction modules for TFDPM. They firstly fuse the features extracted with models like GAT and TCN and then send the fused features to GRU to generate $\boldsymbol{F}_t$. As a comparison, we also GRU for feature extraction which directly obtains $\boldsymbol{F}_t$ based on the input data. Table \ref{table 1} shows the best F1 scores, corresponding precision and recall for each method on all datasets. The best two results for each dataset are shown in bold. 

Isolation forest performs very well on SWAT which has minimum number of sensors. However, we find that the observed values from sensors are usually more complex than those collected data from actuators in these three datasets. Isolation forest performs worse than some recently proposed deep learning-based methods on the other two datasets which have much more sensors. 

In addition, Sparse-AE performs better than LSTM-AE on all datasets, which indicates that proper latent representations are very important for autoencoders. What's more, deterministic methods like LSTM-AE and Sparse-AE usually perform worse than those probabilistic generative models like DAGMM, OmniAnomaly and USAD. It indicates that modeling the distribution of datasets is more effective and robust than deterministic methods.

Our methods obtain best results on all three datasets, especially on WADI and PUMP. The best F1 score based on TFDPM is basically the same with that of OmniAnomaly on SWAT. We analyze the data in SWAT and find that the states of actuators change frequently and some areas marked as abnormal are very similar to normal data. These characteristics make the recall of all the models relatively low. The best F1-scores on WADI and PUMP based on TFDPM which adopts TCN-GAT for feature extraction are about 4\% and 2\% higher than the best results of existing baselines respectively. In addition, TFDPM based on Double-GAT and GRU also obtain good results on all three datasets. It indicates that the conditional diffusion probabilistic model is very effective and can better model the distribution of normal data compared with variational autoencoder. In addition, TFDPM based on GRU performs worse than TFDPM based on TCN-GAT or Double-GAT. Therefore, using GAT to explicitly model the relationship between different channels of the multivariate dataset is effective. Compared with TFDPM based on Double-GAT, TFDPM based on TCN-GAT usually obtains slightly better results on these three datasets, which indicates that TCN is more suitable for extracting features of different time scales compared with the time-oriented GAT.

\begin{table}[t]
	\centering
	\caption{Comparison of performance metrics of anomaly detection on three CPS datasets.}
	\scalebox{0.85}{
			\begin{tabular}{cccc|ccc|ccc}
					\hline
					\multirow{2}*{Models}&\multicolumn{3}{c}{PUMP}&\multicolumn{3}{c}{WADI}&\multicolumn{3}{c}{SWAT}\\
					& PRE&REC&F1&PRE&REC&F1&PRE&REC&F1\\
					\hline
					Isolation forest&0.977&0.852&0.729&0.826&0.772&0.798&0.975&0.754&0.850\\
					LSTM-AE&0.438&0.796&0.565&0.589&0.887&0.708&0.945&0.620&0.749\\
					Sparse-AE&0.798&0.737&0.766&0.769&0.771&0.770&0.999&0.666&0.799\\
					LSTM-PRED&0.925&0.581&0.714&0.620&0.876&0.726&0.996&0.686&0.812\\
					DAGMM&0.931&0.798&0.859&0.886&0.772&0.825&0.946&0.747&0.835\\
					OmniAnomaly&0.937&0.840&0.886&0.846&0.893&0.869&0.979&0.753&\textbf{0.851}\\
					USAD&0.984&0.682&0.731&0.806&0.879&0.841&0.987&0.740&0.846\\
					\hline
					\makecell[c]{\textbf{TFDPM}\\\textbf{(GRU)}}&0.938&0.831&0.881&0.893&0.865&0.879&0.974&0.741&0.842\\
					\makecell[c]{\textbf{TFDPM}\\\textbf{(TCN-GAT)}}&0.959&0.856&\textbf{0.905}&0.939&0.881&\textbf{0.909}&0.989&0.749&\textbf{0.852}\\
					\makecell[c]{\textbf{TFDPM}\\\textbf{(Double-GAT)}}&0.893&0.906&\textbf{0.899}&0.916&0.878&\textbf{0.897}&0.988&0.742&0.848\\
					\hline
			\end{tabular}}
	\label{table_1}	
\end{table}

\begin{figure}[t]
	\centering
	\includegraphics[width=0.98\textwidth]{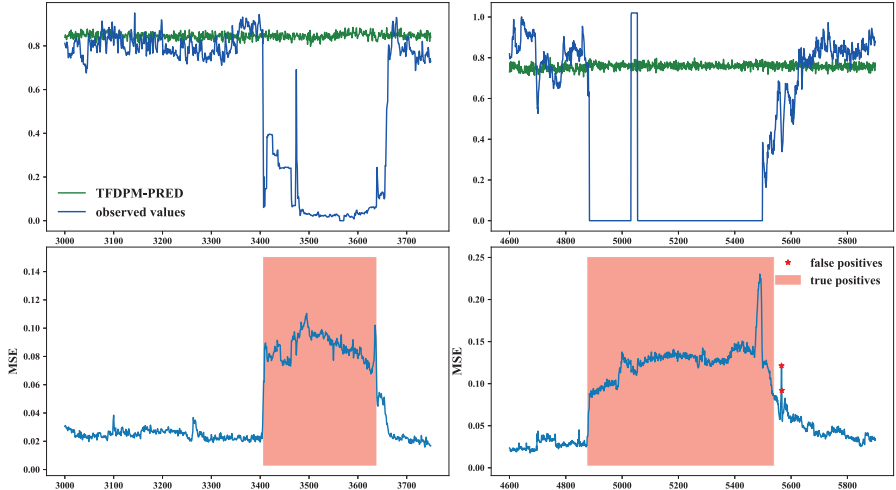}
	\caption{Two examples of anomaly detection based on TFDPM (TCN-GAT).}
	\label{Fig. 7}
\end{figure} 

In order to make the attack detection process more intuitive, two typical anomalous areas in PUMP are shown in Fig. \ref{Fig. 7}. The figures of the first row display observed values and predicted values in the certain channel of the PUMP dataset in different periods. It shows that the observed values drop abruptly over a period of time. Compared with the observed values, the predicted values in this channel remain stable and fluctuate slightly around a certain value. It indicates that TFDPM can well model the distribution of normal data. Even if there are anomalies caused by attacks in some channels, TFDPM can still give reasonable predicted values. As a result, the difference between observed values and predicted values in these areas is larger than those in the normal area. Therefore, we can perform attack detection based on the difference between predicted values and observed values.  After selecting a proper threshold, the anomalous area can be detected. The figures of the second row  confirm this view and they show the mean square error of predictive values and observed values in corresponding periods. The red areas represent correct alarms of TFDPM while the points marked in red represent false positives.  It's obvious that TFDPM can well locate the anomalous areas based on the MSE values.

\begin{figure}[t]
	\centering
	\includegraphics[width=0.9\textwidth]{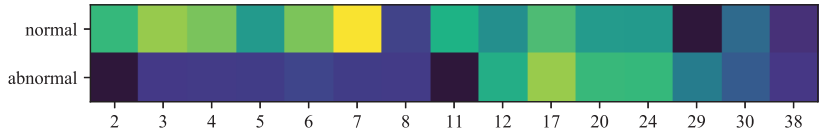}
	\caption{Comparison of attention weights of normal data. Darker color represents lower attention scores.}
	\label{Fig. 8}
\end{figure}

% 异常发生时观测数据的特征往往与正常数据不同，这使得提取的特征具有较大变化。为了直观显示
%The collected data when the system is attacked is usually different with that of 
In order to intuitively show the effect of GAT in TFDPM, part of the attention weights of the feature-oriented GAT in Double-GAT are shown in Fig. \ref{Fig. 8}. It shows the correlation between the data in the first channel and the corresponding channels. Darker color represents lower attention scores. For the data in the same channels under normal and abnormal status, the greater the difference between colors, the correlation between the data of this channel and the first channel changes greater in case of system failure. Therefore, the correlation of the data between certain channels is significantly  different from normal conditions when the system fails. In addition, it indicates that GAT can explicitly model the correlation of the data in different channels, which is extremely important for feature pattern extraction of the data.

\begin{table}[t]
	\centering
	\caption{The performance metrics of TFBDM based on the trained TFDPM models.}
	\scalebox{0.7}{
		\begin{tabular}{ccccc|cccc|cccc}
			\hline
			\multirow{2}*{Models}&\multicolumn{4}{c}{PUMP}&\multicolumn{4}{c}{WADI}&\multicolumn{4}{c}{SWAT}\\
			& PRE&REC&F1&\makecell[c]{Speed\\ratio}&PRE&REC&F1&\makecell[c]{Speed\\ratio}&PRE&REC&F1&\makecell[c]{Speed\\ratio}\\
			\hline
			\makecell[c]{TFBDM\\(TCN-GAT)}&0.958&0.840&\textbf{0.896}&\textbf{3.1}&0.918&0.874&\textbf{0.895}&\textbf{2.8}&0.985&0.757&\textbf{0.856}&\textbf{2.3}\\
			\makecell[c]{TFBDM\\(Double-GAT)}&0.902&0.906&\textbf{0.904}&\textbf{3.2}&0.921&0.878&\textbf{0.899}&\textbf{2.6}&0.984&0.753&\textbf{0.853}&\textbf{2.6}\\
			\hline
	\end{tabular}}
	\label{table_2}	
\end{table}

\subsubsection{Comparison of TFBDM and TFDPM}
Based on the trained TFDPM models in Table \ref{table 1}, corresponding noise scheduling networks are trained based on Algorithm \ref{train_ns}. Then for the extracted feature $\boldsymbol{F}_t$, we construct a sequence of noise vectors and predict values according to Algorithm \ref{t_sampler} at each time step. The results are shown in Table \ref{table_2}.

As shown in Table \ref{table_2}, the best F1 scores of TFBDM are basically the same as those of TFBDM. It's worth noting that TFBDM based on Double-GAT performs slightly better than TFDPM based on Double-GAT on all datasets. In addition, TFBDM based on TCN-GAT also performs better than that of TFDPM based on TCN-GAT on SWAT. Therefore, given the trained TFDPM models, training the noise scheduling network which optimizes the gap between the log-likelihood and the variational lower bound can help further improve the performance of anomaly detection.  What's more, the computation cost of the prediction procedure of TFDPM is large, which makes it unable to meet the requirements of online real-time detection. In order to accelerate the prediction process, we construct shorter noise sequences based on the noise scheduling network. Table \ref{table_2} shows the ratio of the prediction speed of TFBDM to the prediction speed of TFDPM. The results show that the prediction speed of TFBDM can be up to three times that of TFDPM, which greatly reduces the computation cost in the prediction process.

\subsubsection{Influence of hyperparameters}
For TFDPM, the length of historical data $\omega$, the number of diffusion steps $N_s$ and batch size for training are very important. Therefore,  we study the influence of these hyperparameters on the performance of anomaly detection.  

\begin{figure}[t]
	\centering
	\includegraphics[width=0.98\textwidth]{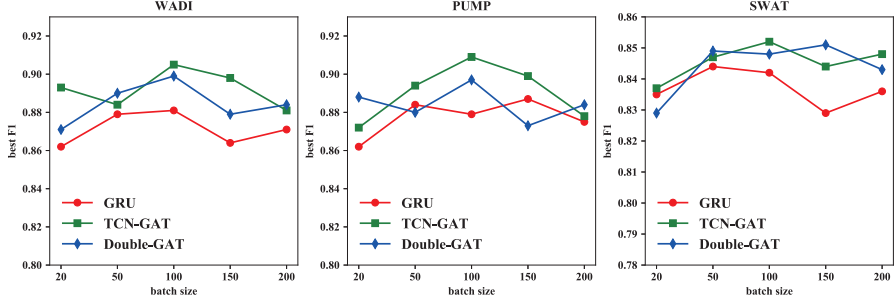}
	\caption{Influence of batch size on the performance.}
	\label{batch}
\end{figure}

\textit{Influence of batch size:} We set the batch size for training as $20, 50, 100, 150, 200$ while keeping all the other parameters unchanged and the results are shown in Fig. \ref{batch}. Firstly, TFDPM based on GRU performs worse than the other models on all the three datasets in most cases. Therefore, compared with directly obtaining $\boldsymbol{F}_t$ with GRU,  we can get better features of historical data by fusing temporal pattern and feature pattern. In addition, the performance of anomaly detection firstly  increases and then decreases with the increase of batch size. For SWAT, TFDPM based on GRU, TCN-GAT and Double-GAT gets their best performance when batch size is set as 50, 100 and 159 respectively. As for the other two datasets, we can get good detection results if the batch size is set as 100. What's more, TFDPM based on TCN-GAT performs slightly better than TFDPM based on Double-GAT on all datasets, which again indicates that TCN is more suitable for temporal pattern extraction than GAT.

\begin{figure}[t]
	\centering
	\includegraphics[width=0.98\textwidth]{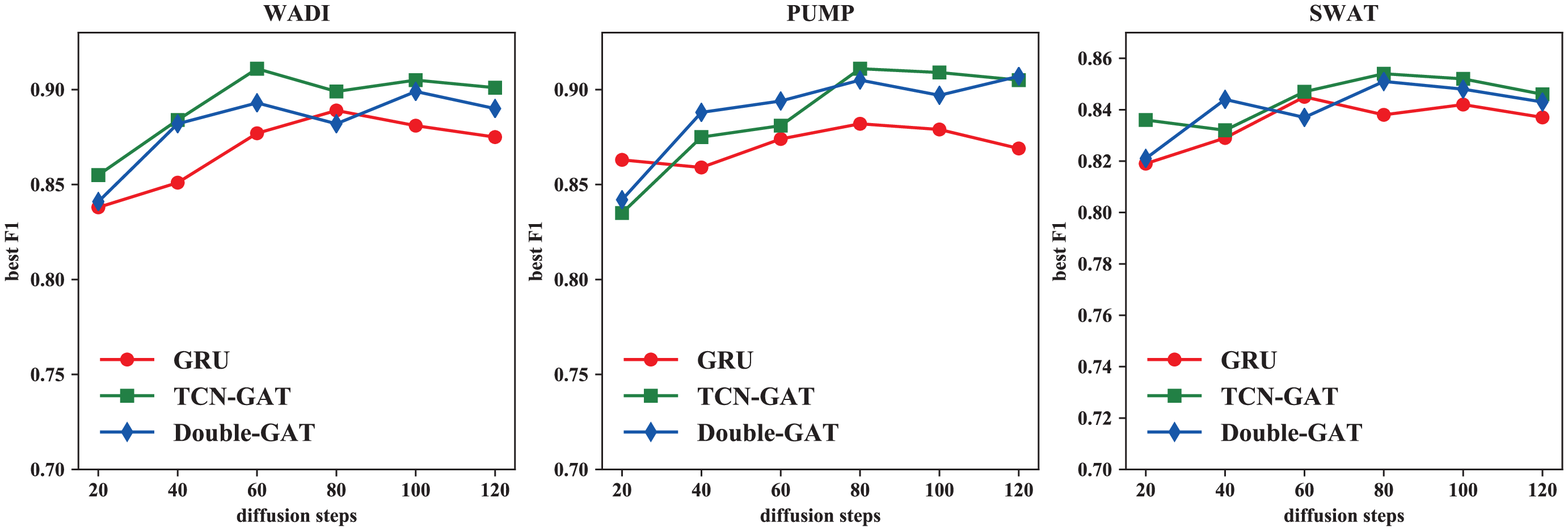}
	\caption{Influence of diffusion steps on the performance.}
	\label{step}
\end{figure}

\textit{Influence of diffusion steps:} The number of diffusion steps is very important for the conditional diffusion probabilistic models. According to Eq. \ref{6}, the target distribution will tend to standard Normal distribution when $N_s$ is big enough. In addition, with the signal-to-noise ratio implemented into the training objective in Eq. \ref{loss}, we can get $\mathbb{L}_t^{2N_s}-\mathcal{L}^{N_s}<0$ \cite{kingma2021variational}. Therefore, large diffusion steps can help further improve the performance. However, large diffusion steps will increase the computational cost. So there is a trade-off between performance and computational cost.  We set diffusion steps as $20,40,60,80,100,120$ while keeping all the other parameters unchanged and the results are shown in Fig. \ref{step}. TFDPM performs better with the increase of the number of diffusion steps, which demonstrate above derivations. However, we can not get better performance when $N_s>100$ if all the parameters are kept unchanged. For the PUMP dataset, TFDPM gets the best results when $N_s=80$.  As for the SWAT dataset, TFDPM based on TCN-GAT and Double-GAT get the best results when $N_s=80$ while $N_s=60$ for TFDPM based on GRU. As for the WADI dataset, TFDPM based TCN-GAT, GRU and Double-GAT get the best results when $N_s=60, 80, 100$ respectively. In addition, TFDPM based on GRU usually performs worse than the other two variants, which again demonstrates that we can get better representations by explicitly modeling the feature pattern. 

\begin{figure}[t]
	\centering
	\includegraphics[width=0.98\textwidth]{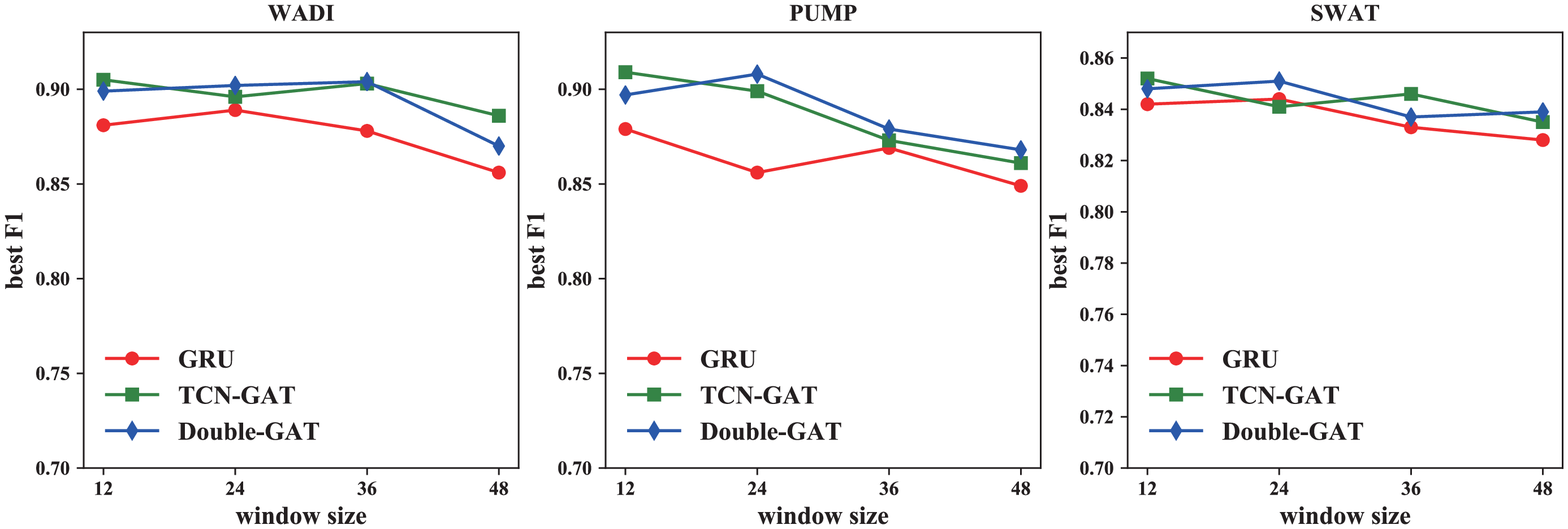}
	\caption{Influence of window size on the performance.}
	\label{window}
\end{figure}

\textit{Influence of window size:} The length of the historical data $\omega$ is crucial for the temporal pattern and feature pattern extraction module. We set the window size as $12, 24, 36, 48$ while keeping all the other parameters unchanged and the results are shown in Fig. \ref{window}. The best F1 scores of TFDPM based on TCN-GAT and Double-GAT on WADI fluctuate in a small range when $\omega$ changes from 12 to 36. But the performance of these two models on WADI decreases when $\omega=48$. As for the other datasets, the performance of TFDPM usually decreases when $\omega$ is greater than 24. Therefore, longer sequences may not be helpful for feature extraction of temporal pattern and feature pattern while keeping all other parameters unchanged. What's more, the performance of TFDPM based on GRU usually performs worse than the other variants, which again indicates that we can get better representations of historical data by fusing the features extracted from temporal and feature pattern. 

In summary, we explore the influence of hyperparameters on these three datasets. The performance of TFDPM based on TCN-GAT is generally slightly better than TFDPM based on Double-GAT. In addition, TFDPM based on GRU usually performs worse than the other two models. The reasonable values for batch size, the number of diffusion steps and the length of window size can be set as 100, 100, 12 for all datasets.

\section{Conclusion and future works}\label{conclusion}
In this paper, we propose TFDPM, a general framework for attack detection tasks in CPSs. It consists of two components: temporal pattern and feature pattern extraction module and conditional diffusion probabilistic module. Particularly, we use graph attention networks to explicitly model the correlation of data in different channels in the first module. In addition, the energy-based generative model used in the second module is less restrictive on functional forms of the data distribution. To realize real-time detection, a noise scheduling network is proposed for accelerating the prediction process. Experiments show that TFDPM outperforms existing state-of-the-art methods and the noise scheduling network is efficient.

There is still much work to be done. For example, discrete flows \cite{tran2019discrete} can be used to model discrete signals in collected data. Besides, the conditional diffusion probabilistic model used in this paper can be view as the discrete form of stochastic differential equations (SDEs) \cite{song2020score}. Therefore, more powerful generative models can be designed based on SDEs.

%TODO: 本文的结论，未来工作中如何加速，如何替换模块，针对离散数据可以采用离散随机流模型； 本文提出了，，，一种框架，分为xx两部分。其中为显式建模，，，采用，，；此外，采用基于xx生成哦模型，约束较少；为满足实时性的要求，采用，，，，实验结果表明，，，gat作用，dpm作用，加速作用；
% 未来工作：对于离散信号可以采用离散的diffusion models解决；将其纳入更广泛的SDE框架中考虑，如何使得方差最小，变化尽量快

%TODO: Appendix: denoising score matching的推导； 关于conditional noise scheduling network中的结论推导

%\section*{References}
\appendix
\section{}\label{A}
\begin{equation} \label{A1}
\begin{aligned}
\nabla_x\textbf{s}_{\theta}(\boldsymbol{x}_t^n, \alpha_n \boldsymbol{F}_t)&=\nabla_{x}^2\log p_{\theta}(\boldsymbol{x}_t^{n-1}|\boldsymbol{x}_{t}^{n}, \boldsymbol{F}_t)\\
&=\nabla_x \big[-\frac{1}{2}\log 2\pi -\log \sqrt{\Sigma_{\theta}\textbf{I}}-\frac{(\boldsymbol{x}_t^n - \tilde{\boldsymbol{\mu}}_{\theta}(\boldsymbol{x}_t^n, \alpha_n, \boldsymbol{F}_t))^2}{\Sigma_{\theta}\textbf{I}}\big]\\
&= -\frac{\boldsymbol{x}_t^n - \tilde{\boldsymbol{\mu}}_{\theta}(\boldsymbol{x}_t^n, \alpha_n, \boldsymbol{F}_t)}{\Sigma_{\theta}\textbf{I}}\\
\end{aligned}
\end{equation}

In addition, according to the definition of $\epsilon_{\theta}$, we can get

\begin{equation}\label{A2}
	\boldsymbol{\epsilon}_{\theta}(\boldsymbol{x}_t^n, \alpha_n, \boldsymbol{F}_t) = \frac{(\boldsymbol{x}_t^n - \boldsymbol{\mu}_{\theta}(\boldsymbol{x}_t^n, \alpha_n, \boldsymbol{F}_t))}{\sqrt{\Sigma_{\theta}\textbf{I}}}.
\end{equation}
Combining Eq. \ref{A1} and Eq. \ref{A2}, we can get $\boldsymbol{\epsilon}_{\theta}(\boldsymbol{x}_t^n, \alpha_n, \boldsymbol{F}_t) = -\sqrt{\Sigma_{\theta}\textbf{I}}\textbf{s}_{\theta}(\boldsymbol{x}_t^n, \alpha_n \boldsymbol{F}_t)$. If we set  $\Sigma_{\theta}=\tilde{\beta}_n \textbf{I}$ , then with similar derivation, Eq. \ref{loss} can be written as 
\begin{equation}
	\begin{aligned}
		\mathcal{L}_t&= \mathbb{E}_{\boldsymbol{x}_t^0, \boldsymbol{\epsilon}, n}\big[\frac{N_s}{2}(\text{SNR}(n-1)-\text{SNR}(n))\Vert\boldsymbol{\epsilon} - \boldsymbol{\epsilon}_{\theta}(\boldsymbol{x}_t^n, \alpha_n, \boldsymbol{F}_t)\Vert^2\big]\\
		&= \mathbb{E}_{\boldsymbol{x}_t^0, \boldsymbol{\epsilon}, n}\big[\frac{N_s}{2}(\text{SNR}(n-1)-\text{SNR}(n))\Sigma_{\theta}\textbf{I}\Vert\boldsymbol{s} - \boldsymbol{s}_{\theta}(\boldsymbol{x}_t^n, \alpha_n, \boldsymbol{F}_t)\Vert^2\big].
	\end{aligned}
\end{equation}

Therefore, the training objective in Eq. \ref{loss} is a weighted version of denoising score matching objective.

\section{}\label{B}
For the trained TFDPM, assume $p_{\theta^*}(\boldsymbol{x}_t^{1:n-1}|\boldsymbol{x}_t^0, \boldsymbol{F}_t)=q(\boldsymbol{x}_t^{1:n-1}|\boldsymbol{x}_t^0, \boldsymbol{F}_t)$, then the gap between the log-likelihood $\log p(\boldsymbol{x}_t^0|\boldsymbol{F}_t)$ and the variational lower bound $F_{\text{score}}^{t, n}(\theta^*)$ can be formulated as
\begin{equation}
	\begin{aligned}
&\log p(\boldsymbol{x}_t^0|\boldsymbol{F}_t) - F_{\text{score}}^{t, n}(\theta^*)\\ 
&=\log p_{\theta}(\boldsymbol{x}_t^0|\boldsymbol{F}_t) - \mathbb{E}_{q(\boldsymbol{x}_t^n|\boldsymbol{x}_t^0)}\big[\mathbb{E}_{p_{\theta^*}(\boldsymbol{x}_t^{1:n-1}|\boldsymbol{x}_t^n)}\big[\log \frac{p_{\theta^*}(\boldsymbol{x}_{t}^{0:n-1},\boldsymbol{F}_t)}{p_{\theta^*}(\boldsymbol{x}_{t}^{1:n-1}|\boldsymbol{x}_t^n, \boldsymbol{F}_t)}\big]\big]\\
&=\mathbb{E}_{q(\boldsymbol{x}_t^n|\boldsymbol{x}_t^0)}\big[\mathbb{E}_{p_{\theta^*}(\boldsymbol{x}_t^{1:n-1}|\boldsymbol{x}_t^n)}\big[\log\frac{p_{\theta^*}(\boldsymbol{x}_t^{1:n-1}|\boldsymbol{x}_t^n,\boldsymbol{F}_t)}{p_{\theta^*}(\boldsymbol{x}_{t}^{1:n-1}|\boldsymbol{x}_t^0,\boldsymbol{F}_t)}\big]\big]\\
&=\mathbb{E}_{q(\boldsymbol{x}_t^n|\boldsymbol{x}_t^0)}\big[\mathbb{E}_{p_{\theta^*}(\boldsymbol{x}_t^{1:n-1}|\boldsymbol{x}_t^n)}\big[\log\frac{p_{\theta^*}(\boldsymbol{x}_t^{1:n-1}|\boldsymbol{x}_t^n,\boldsymbol{F}_t)}{q(\boldsymbol{x}_{t}^{1:n-1}|\boldsymbol{x}_t^0,\boldsymbol{F}_t)}\big]\big]\\
&=\mathbb{E}_{q(\boldsymbol{x}_t^n|\boldsymbol{x}_t^0)}\big[\sum_{i=2}^nD_{\text{KL}}(p_{\theta^*}(\boldsymbol{x}_t^{i-1}|\boldsymbol{x}_t^i, \boldsymbol{F}_t)\Vert q_{\phi}(\boldsymbol{x}_t^{i-1}|\boldsymbol{x}_t^0, \boldsymbol{F}_t))\big]\\
&=\mathbb{E}_{q(\boldsymbol{x}_t^n|\boldsymbol{x}_t^0)}\big[\sum^n_{i=2}\mathcal{L}(\phi; \theta^*)\big]\\
	\end{aligned}
\end{equation}
The last equality assumes that $q_{\phi}(\boldsymbol{x}_t^{n}|\boldsymbol{x}_t^{n-1}) = q(\boldsymbol{x}_t^{n}|\boldsymbol{x}_t^{n-1})$ when $\tau=1$.
%TODO: score matching等价性； conditional BDMM推导

\bibliography{mybibfile}

\end{document}